\documentclass[journal]{IEEEtran}
\pdfoutput=1
\usepackage{multirow}
\usepackage{subcaption}
\usepackage{color}
\usepackage[]{algorithm2e}
\usepackage{verbatim}
\usepackage{amsmath,amssymb,amsfonts,amsthm}
\usepackage{atbegshi} 
\AtBeginDocument{\AtBeginShipoutNext{\AtBeginShipoutDiscard}}
\usepackage{balance}
\usepackage{url}

\ifCLASSINFOpdf
  \usepackage[pdftex]{graphicx}
\else
\fi
\begin{document}
%
\title{A Multi-Stage Algorithm for Acoustic Physical Model Parameters Estimation}
%
%
%

\author{Leonardo Gabrielli, Stefano Tomassetti, Stefano Squartini, Carlo Zinato, Stefano Guaiana}
\thanks{The first three authors are with the Dept. Information Engineering, Universit\`a Politecnica delle Marche, the last two are with Viscount International SpA}
\thanks{}
\thanks{Manuscript received ---; revised ---.}

%
%

\markboth{}
{name \MakeLowercase{\textit{et al.}}: title}
%



\maketitle

\begin{abstract}
One of the challenges in computational acoustics is the identification of models that can simulate and predict the physical behavior of a system generating an acoustic signal. Whenever such models are used for commercial applications an additional constraint is the time-to-market, making automation of the sound design process desirable. In previous works, a computational sound design approach has been proposed for the parameter estimation problem involving timbre matching by deep learning, which was applied to the synthesis of pipe organ tones. In this work we refine previous results by introducing the former approach in a multi-stage algorithm that also adds heuristics and a stochastic optimization method operating on objective cost functions based on psychoacoustics. The optimization method shows to be able to refine the first estimate given by the deep learning approach and substantially improve the objective metrics, with the additional benefit of reducing the sound design process time. Subjective listening tests are also conducted to gather additional insights on the results.
\end{abstract}


%

\pagenumbering{arabic}

\section{Introduction}
\label{sec:intro}

In the recent years a number of advances have been done in several fields of computational acoustics, from the emulation of 3D spaces by finite difference modelling \cite{Hamilton2017FDTD} to nonlinear strings \cite{Desvages2016PhyMod} and plates \cite{Torin2014nonlinear}. { Among well-known methods \cite{Valimaki2004review,Valimaki2005review}, some rely on an accurate description of a physical setting, and its parameters, thus, can be measured from real settings or purportedly generated to simulate a specific one. Other popular techniques, such as those based on digital waveguide modelling \cite{Smith1992CMJ,KarjValTolo98,Smith2002DWG}, modal synthesis \cite{Adrien1991Modal,Bank10:ASL,Trautmann03:Book,Zambon2013expressive} or a mix of the aforementioned methods \cite{Bensa2003,Karjalainen2004,Lee10:ASL,GabrielliFDTDAES2013}, impose simplifying hypotheses that produce lower-complexity solutions and, hence, a lower computational cost overall. This is done, however, at the cost of a departure from the underlying physics, thus, requiring strategies to estimate coefficients and parameters by matching the outcomes of the model and the target. In other words, a \textit{timbre matching} algorithm is required to provide an optimal solution, at least in psychoacoustic terms}.

{In the recent years, several works suggested solutions to estimate some of the parameters using statistical and numerical techniques \cite{Chatziioannou2012estimation,Wilkinson2017LatentModal,Taillard2018Modal,Kobayashi2018KautzPiano}. While these estimation techniques are based on prior knowledge of the model and in some cases are aimed to find only a subset of the parameters, a black-box approach could model the entire system without prior physical knowledge and with the added benefit of being easily applied to different models.} The first known attempt to calibrate a physical model using such an approach is reported in \cite{CemgilErkut1997Calibration}, where a multilayer perceptron (MLP) \cite{Rumelhart86-LRB} is trained to estimate parameters of a simple Karplus-Strong \cite{Karplus1983} model by learning a perceptual distance obtained from subjective listening tests. Recent works \cite{Gabrielli2017DAXF, Gabrielli2017TETCI}, extend the topic by introducing additional concepts and techniques, formalizing the goals and problems of the \textit{timbre matching} task in a \textit{computational sound design} setting. More specifically, state of the art end-to-end learning \cite{Goodfellow2016Book} has been used with the goal of matching a desired timbre. Convolutional Neural Networks (CNN) have been proposed, which are able to learn acoustic features and provide estimates of the parameters of a flue pipe physical model \cite{Zinato2008patent} and the extended Karplus-Strong model. Data visualization techniques such as T-SNE \cite{Maaten2008TSNE} are proposed for dataset exploration as well as objective metrics for evaluating the performance of the estimation algorithm.

{Other works that are strictly related to computational sound design are reported in \cite{Mitchell2007evolutionary,Mitchell2012evolutionary,Tatar2016OP1}, where genetic algorithms are used to estimate coefficients of synthesizers.} Specifically, in \cite{Mitchell2007evolutionary} the concept of contrived and non-contrived matching is introduced, which is used also in this work. \textit{Contrived tones} are those belonging to the space of all the signals that can be generated by the synthesis engine, while \textit{non-contrived tones} come from any other sound source (either virtual or real). The paper evaluates results in terms of spectral Euclidean distance. The envelope Euclidean distance is proposed in  \cite{Tatar2016OP1} and the relative spectral error in  \cite{Mitchell2012evolutionary}. In \cite{Itoyama2014Estimation} linear regression is used to estimate parameters using the Source-to-Distortion Ratio (SDR) \cite{Vincent2006BSS}, however the definition is not clear and there is no detail on how this can be applied to the regression scenario described in the paper.

Empirical evaluations are conducted in \cite{Tatar2016OP1} with sound designers, showing that the sound design accuracy of the proposed algorithm is superior to the one obtained by humans. The work reports, however, that the time required for timbre matching by means of the proposed approach is longer than the time required by a sound designer. The reduction of the time to complete the sound design process should be an additional goal for computational sound design processes to be practically relevant {and gives further motivation to this work}.

{To the best of our knowledge, the works in \cite{Mitchell2007evolutionary,Mitchell2012evolutionary,Tatar2016OP1,Itoyama2014Estimation} are the only to propose a black-box approach for the parameter estimation task in the virtual instruments literature, however the comparison with these experiments is not straightforward due to differences in the metrics and the synthesis engines.}


In the present work we expand in many respects the algorithms detailed in \cite{Gabrielli2017DAXF, Gabrielli2017TETCI}, motivated by the need to improve the achieved performance. {We apply the novel technique to organ pipe tones to compare with our previous works, and extend the algorithm with the introduction of a optimization procedure that refines the search of a solution.} This, in turn, requires introducing metrics in the estimation framework. The evaluation now is performed on a larger number of target sounds with different characteristics. Subjective listening tests are conducted to add further insights to the objective evaluations. {Finally, we conducted an informal study on the sound design process before and after the introduction of the proposed method, to conclude on its impact in terms of time and usability.} 

The outline of the paper follows. We first describe the proposed method in all its parts in Section \ref{sec:approach}. Then in Section \ref{sec:physis} we introduce the use case for this paper, the dataset and specific adaptations related to the use case, such as features and metrics. In Section \ref{sec:results} we describe the experimental setting and report objective and subjective results. The conclusions are drawn in Section \ref{sec:conclusion}.


\section{Proposed Method}
\label{sec:approach}

\subsection{Overview}

The parameter estimation problem is intended along this paper as the problem of finding a set of parameters $\boldsymbol{\theta}$ that allows a physical model $f(\boldsymbol{\theta})$ to produce an output signal $\hat{s}[n] = f( \boldsymbol{\theta} )$  that matches a target signal $s[n]$ as close as possible. The matching needs to be measured in psychoacoustic terms, as the physical model generally differs from the actual mechanism that produces $s[n]$, thus, a perfect samplewise match cannot be expected. In \cite{Gabrielli2017TETCI} an approach employing a CNN architecture was devised with the goal of estimating $\boldsymbol{\theta}$ from $s[n]$ given prior knowledge of the model {in terms of the tuples $(\boldsymbol{\theta}, s[n])$. This corpus of data was generated by the model itself and employed in a supervised fashion for training.}

\begin{figure*}[htbp]
\centering
\includegraphics[width=1\linewidth]{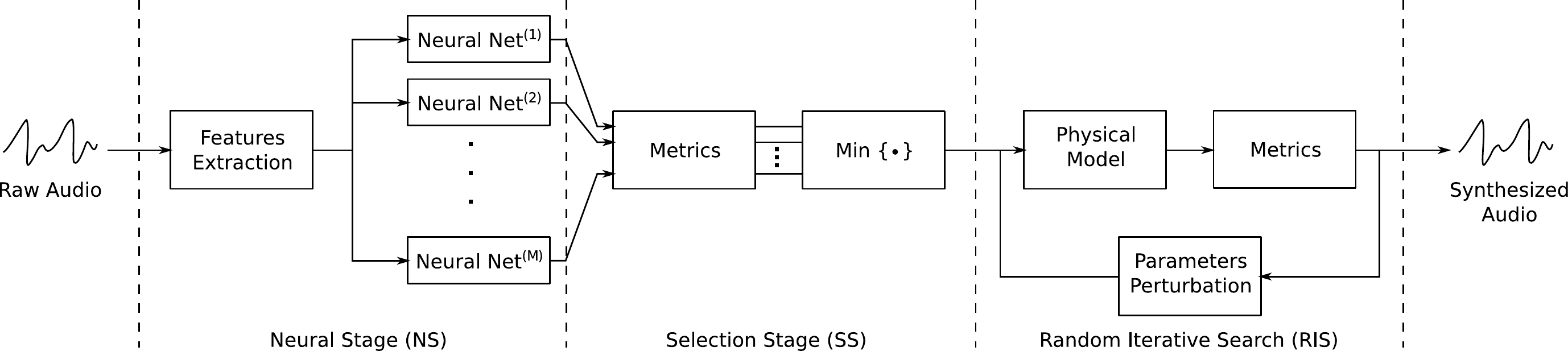}
\caption{Multi-Stage Algorithm Overview. Neural Stage (NS) and Selection Stage (SS) are part of a \textit{global search}, the Random Iterative Search (RIS) composes the \textit{local search}.}
\label{fig:NN_SA_HSRA}
\end{figure*}

In this work we propose a multi-stage approach consisting of a Neural Stage (NS), a Selection Stage (SS) and a metaheuristic stage based on a Random Iterative Search (RIS) algorithm, as shown in Figure \ref{fig:NN_SA_HSRA}. The NS is composed of fully-connected neural networks that provide different estimations given the input features of the target signal. The SS evaluates each of these estimates and selects, for each note, the best ones, based on one or more acoustic metrics.

The first two stages are meant to perform a \textit{global search}, thus, providing a first solution that can still be improved upon with a local search. For this goal we introduce a general stochastic optimization method, the RIS and an extension thereof named Multi-Objective RIS (MORIS), that looks for a local minimum by iterative stochastic perturbation of the parameters. The motivation behind the introduction of an optimization algorithm is the large acoustical error introduced by small estimation errors for some physical parameters. The optimization stage is expected to refine the results by reducing the estimation errors on acoustical basis. Both the SS and the MORIS employ acoustic metrics, later described in Section \ref{subsec:obj-metrics}.

\subsection{Neural Stage}
\label{subsec:neural-networks}

Convolutional neural networks were first shown to be capable of estimating acoustic physical model parameters in \cite{Gabrielli2017DAXF}. One advantage of CNN is the autonomous learning of the input features at the additional expense of large training and testing times. {In principle, no human intervention is required, as the training process guarantees the features that minimize the cost function on a given dataset. Features learning allows CNN to be applied to many different use cases. Furthermore, a successfully trained CNN can often be repurposed, with the burden of a partial retraining \cite{Pratt2012Transfer}, to adapt it to the new data. This is called transfer learning.} Unfortunately these advantages may be vanished by the need for a lengthy hyperparameter search. A large number of network models are trained, each with different hyperparameters, including the optimizer, the cost function, the size of the short time Fourier transform (STFT) frames, the kernels and their strides, the layers and their number, etc. Regularization techniques are usually required to avoid overfitting, that prevents a network from generalizing to unseen data. No hyperparameter search method can guarantee the optimal solution, thus only a thorough random search can yield good results, at the expense of computational times.

Differently from other machine learning application fields, in this application field we can rely on basic prior knowledge of the data. If this knowledge is applied to the feature extraction phase, efficient digital signal processing can be used to hand-craft features in place of the convolutional layers. This reduces the computational cost but we argue that it also increases the accuracy of the input features. The use of Logmel or STFT spectra with low frequency resolution, that are common to deep learning audio applications, cannot provide accurate information about partials frequency and amplitudes. Furthermore, the max-pooling layers in a CNN although valuable for information compression, reduce the position accuracy of the activation peaks \cite{Sabour2017capsules}, that in the present case may be related to partials, and, thus, may lead to a loss of vital information for timbre matching and other auditory processing tasks.

For this reason we propose the use of hand-crafted features based on a model of the input signal that may provide more accurate information to the neural network. As an example, if adopting a sinusoidal plus noise model, the tone will be described by a deterministic and a stochastic component \cite{Serra1997SinePlusNoise}, yielding features such as the partials frequency, and the evolution of their amplitudes with time, the evolution of the noise energy and the initial ratio between the deterministic and stochastic components. The neural network will, thus, map these data to the model parameters. Since the hand-crafting of the features depends on the input signal model it will be discussed in Section \ref{subsection:features} for the application case proposed afterward. 

By stripping the convolutional layers from a CNN we retain the fully-connected layers only, what is generally called a multilayer perceptron (MLP). The gained computational resources can be, thus, employed to implement $M$ parallel neural models, each one trained with different data to take advantage of the diversity of multiple alternative estimates. These estimates shall be candidates for the selection stage (SS), discussed in Section \ref{subsec:selection-algo}.

\subsection{Selection Stage}
\label{subsec:selection-algo}

The selection stage is based on the assumption that $M$ estimates are provided by $M$ neural networks from one input signal. Since the neural networks are all trained differently, some of their estimates could be better than others. Ideally, the SS is an automation of a manual selection conducted by a human expert that is, however, expensive in terms of time and does not guarantee a unique solution.

In our previous works we remarked that the acoustic error perceived by listeners has no straightforward connection to the error on physical model parameters, due to the different role and weight of each parameter in shaping the output signal. For this reason, together with the parameters error, some acoustic metrics were proposed in \cite{Gabrielli2017DAXF,Gabrielli2017TETCI} to evaluate the results. 
The use of metrics, however, can be successfully employed to automate the selection of the candidates from the previous stage and, thus, improve the algorithm performance without human intervention. The metrics are, thus, employed to select the best candidate among the ones provided in the NS. Prior knowledge of the data is required to devise the metrics. Section \ref{subsec:obj-metrics} will report the ones devised for the use case of Section \ref{sec:physis}.

The selection is based on the iterative comparison described by Algorithm \ref{algo:SS} which selects the best matching tone for a given target tone according to the acoustic metrics. Please note that for producing a whole range of tones (e.g. notes on the keyboard range) the proposed algorithm should be repeated for each single tone. 
\begin{algorithm}
\KwData{\{$\boldsymbol{\theta_1}, ... \boldsymbol{\theta_M}$\}}
$\boldsymbol{\theta_{b}}$ := $\boldsymbol{\theta_1}$\;
\For {$i$ in $M$}{
compute $\hat{s}_i[n] = f(\boldsymbol{\theta_i})$\;
evaluate $d_i = \sum_{k=1}^{K} a_k J_k(s[n], \hat{s}_i[n])$\;
\If{$d_i < d_{i-1}$}{$\boldsymbol{\theta_{b}}$ := $\boldsymbol{\theta_i}$}
}
return $\boldsymbol{\theta_{b}}$\;
\caption{Selection Algorithm. The best estimate $\boldsymbol{\theta_{b}}$ is computed using a mix of $L$ metrics weighted by coefficients $a_k$.}
\label{algo:SS}
\end{algorithm}

At the output of the selection stage there is a dimensionality reduction of $1/M$ where $M$ is the number of candidates, i.e. of neural networks employed in the NS.
The metrics are computed on the signals generated by each one of the candidates. Please note that this requires generating a signal for each candidate by running the physical model, therefore, the computational cost of this stage depends on the cost of the physical model. The weighting of the metrics $a_k$ is arbitrary and can be imposed by a sound designer supervising the computational sound design process.



\subsection{Random Iterative Search}
\label{subsec:opt-algo}

The outcome of the NS and the SS is not guaranteed to be optimal. The RIS optimization algorithm is intended as a refinement of the estimate that operates iteratively based on the minimization of one acoustic metric (RIS) or multiple acoustic metrics (MORIS), providing a local search of a distance minimum that starts from the estimate at the output of the SS. A similar approach to the problem can be found in \cite{Mitchell2007evolutionary,Mitchell2012evolutionary,Tatar2016OP1}, where evolutionary algorithms are exploited for parameter search. In this case, however, the starting point is not selected randomly but is the outcome of a global search operation that is expected to speed up the convergence of the iterative search.

The basic RIS algorithm and its extension, MORIS, are based on the random perturbation of the parameter space, weighted by the distance calculated from the metrics. The perturbed solution is discarded if its distance from the target signal has increased with respect to the previous step. The entity of the random perturbation, thus, decreases as the algorithm approaches a match. The random perturbation of the parameters at step $i$ is done according to the following equation:

\begin{equation}
\boldsymbol{\theta_i} = \mu d_i \cdot (\boldsymbol{\theta_{b}} \circ [ \mathbf{r} \circ \mathbf{g}])
\label{eq:RIS}
\end{equation}
where: 
\begin{itemize}
\item $ \boldsymbol{\theta_{b}} $ is the best parameter vector achieved so far,
\item $\mu < 1$ is an arbitrary step size fixed by the user to improve convergence,
\item $d_i$ is the distance at step $i$,
\item $ \mathbf{r} $ is a sparse vector of random values $\in {0, 1}$ of the same size as $ \boldsymbol{\theta_{b}} $,
\item $ \mathbf{g} $ is the perturbation vector, with values following a Gaussian distribution and having same size as $ \boldsymbol{\theta_{b}} $.
\item $\circ$ is the element-wise Hadamard product operator
\end{itemize}
{The use of a sparse vector $ \mathbf{r} $ allows only a random subset of the parameters to be perturbed at each iteration, reducing the dimensionality of the problem. In essence, only a subset of the parameter vector $ \boldsymbol{\theta_{b}} $ is perturbed at each step, with random values weighted by the distance, to scale down the step size during convergence.}

The difference between RIS and MORIS is in the evaluation of the cost function. While the RIS has a cost function corresponding to a single metric, with the MORIS algorithm, the cost function is composed by a weighted sum of $K$ metrics, i.e.
\begin{equation}
d_i = \sum_{k=1}^{K} b_k J_k(s[n], \hat{s_i}[n])
\end{equation}
where the weights $b_k$ may be different from the weights $a_k$ seen in the SS. The weighting is arbitrary and may be determined to favor the matching of some characteristics of the signal with respect to others.

The general algorithm including both RIS and MORIS variants is reported as Algorithm \ref{algo:RIS}.
\begin{algorithm}
compute $\hat{s}_0[n] = f( \boldsymbol{\theta_0} )$\;
evaluate $d_0 = \sum_{k} b_k J_k(s[n], \hat{s}_0[n])$\;
\While{$d_i < \epsilon$ \textbf{OR} maximum iteration reached \textbf{OR} $p$ iterations reached}{
$\boldsymbol{\theta_i}$ := random perturbation of $\boldsymbol{\theta_{b}}$ weighted by $d_{b}$\;
compute $\hat{s}_i[n] = f( \boldsymbol{\theta_i} )$\;
evaluate $d_i = \sum_{k} b_k J_k(s[n], \hat{s}_i[n])$\;
\If{$d_i < d_{i-1}$}{
$\boldsymbol{\theta_{b}}$ := $\boldsymbol{\theta_i}$\;
$d_{b}$ := $d_i$
}
}
\caption{Multi-Objective Random Iterative Search. The algorithm is based on a random perturbation weigthed by the distance of the last best step $d_b$. It must be noted that the random perturbation is sparse, i.e. not all parameters are perturbed at a given iteration, as reported in Equation (\ref{eq:RIS}). Please note that with $K=1$ the algorithm reduces to RIS.}
\label{algo:RIS}
\end{algorithm}

{Both RIS and MORIS work with normalized parameters. Before feeding the physical model with the perturbed parameters, denormalization takes place. Normalization consists in a parameters values adaptation in range $[-1, 1]$ done exploiting parameter-wise value range. Denormalization is the inverse procedure. This process is crucial to make the perturbation entity coherent across all dimensions of the parameters vector.
The RIS and MORIS algorithms stop whenever the matching goal is reached, they get stuck in a minimum (early stopping) or the time has expired. Specifically, the following criteria are employed:
\begin{itemize}
\item the distance gets lower than the \textit{distance threshold} $\epsilon $;
\item more than $p$ \textit{patience iterations} (iterations without an improvement) have passed;
\item \textit{maximum iterations} are reached.
\end{itemize}
}
Please note that multiple RIS optimization runs can be performed by applying different cost functions each time, seeking minimization of a specific acoustical aspect for each run. Similarly, a single MORIS run can be performed using a cost function composed of a weighted sum of metrics.


\section{Application to a Pipe Organ Physical Model}
\label{sec:physis}

\subsection{Flue Pipes}
\label{subsec:fluePipes}

A pipe organ generally consists of one or more \textit{manuals} (or \textit{divisions}) and set of \textit{stops}, i.e. groups of pipes. Pipes from a stop share similar construction characteristics such as materials and shape, and thus, provide similar timbre and sonic qualities. One stop may be assigned to one or more manuals from the organ console and can be switched on and off during a performance, allowing changes in dynamics and timbre during the execution of a piece, obtained by summing the stops. This partially addresses the lack of expressivity of the valve opening mechanism that allows no user interaction and is basically dyadic (valve open/closed). The sound design process for the pipe organ can be reduced to the timbre matching of single stops, thus, we shall concentrate on these. 

Most stops have one pipe for each note so the experiments reported in Section \ref{sec:results} were made on this kind of stops. Pipes in a stop are tuned as multiple of a base pitch, which is labelled in terms of the length of the longest pipe in the stop, corresponding to a low C. This length is expressed in feet and is approximately 8 feet (8') for a unison stop. Stops may range from 32', i.e. 2 octaves lower than the unison stops, to 1/2', i.e. 4 octaves higher than the \textit{unison} stops. {Stops can be pitched also at non-octave harmonic intervals, e.g. a pipe of a $2^{2/3}$ stop is shifted by 19 semitones with respect to a unison stop.}

Stops are divided in families depending on their construction features. Flue pipe stops are divided in three families: open stops, closed stops and harmonic stops \cite{AudsleyOrgan}. The term open and closed refers to the termination of the pipe, which can be open or closed, determining the termination impedance of the air column and, thus, the relation between pitch and pipe length and the harmonic content, with closed stops having no even harmonics. Harmonic stops pipes can be either open or closed. They feature a hole along the bore that determines the pitch.

Each family can be divided in subfamilies depending on tonal or construction characteristics. In this work we take the most common subfamily for each one of the families, namely \textit{Principale} for the open stops, \textit{Bordone} for the closed stops and \textit{Flauto Armonico} from the harmonic stops \footnote{Please note that these are the names in the Italian organ building tradition. Stop names reported in this work can slightly vary depending on the country of origin (e.g. the French \textit{Flute Harmonique} for Flauto Armonico or \textit{Bourdon} for Bordone).}.

The Principale stops are richer in frequency components than Bordone stops because the latter are closed pipe stops, typically made in wood. Bordone tones are almost exclusively composed of odd harmonic components and have typically darker and softer timbre. Flauto Armonico stops can sometimes be composed of closed pipes for the lower part of the keyboard. The presence of a hole and the doubled length yields a sub-harmonic at half the fundamental frequency. These stops often have intermediate qualities and sit in between the open and closed stops.

\subsection{Flue Pipe Modelling}
\label{subsec:fluemodel}
The physical model considered here \cite{Zinato2008patent} is meant to simulate a generic flue pipe by means of digital waveguide (DWG) principles. It has been implemented as a standalone C application, allowing it to be run iteratively for use with the proposed algorithm and optimizations have been performed to make the iterations in the MORIS as short as possible. 

{The algorithm is composed of three parts, the harmonic generation, the noise generation and the passive resonator.

The harmonic generation algorithm stems from observation of feedback models that include coupling between a nonlinear wind jet excitation mechanism and the passive resonating bore, as in \cite{Valimaki1992Flute,Hanninen1996Flute}. A signal-wise model of the excitation has been adopted to optimize the algorithm. The pipe model is, thus, feedforward, with a harmonic stimulus fed to a digital waveguide bore model. The stimulus is generated from a sine oscillator at the fundamental frequency. The second harmonic is generated by squaring the fundamental sine wave and removing the offset. These two are independently processed by envelope generators and clipping nonlinearities with individual gains to have a somewhat independent control on even and odd harmonics. Then they are summed and the cascade of a feedforward comb filter and a static nonlinearity is applied to shape the signal similarly to \cite{Valimaki1992Flute}, where a short delay line emulate jet delay and a sigmoid function emulates the nonlinear interaction between the air jet and the pipe tube. Finally a bandpass filter with dry/wet controls centered at the fundamental frequency is applied to remove part of the harmonic components introduced by the nonlinearities described thus far.

In flue pipes, noise is generated by air hitting the pipe lips. In the model this is simulated by a dedicated computational block interacting with the tonal generation and subject to the resonating cavity. The output is obtained by a pink noise filtered by a feedback delay network which comprises a nonlinear clipping function with thresholds proportional to a rate signal, obtained from high pass filtering and rectifying the fundamental frequency sine tone. This gives so-called \textit{noise granulation}, i.e. the periodically pulsed noise typical of flue pipes. An aleatory signal determines slight changes in the pipe pitch to emulate attack turbulence.

The passive resonator models the pipe bore by means of a digital waveguide with frequency-dependent losses by lowpass filtering and a DC blocking filter. The pipe has a dispersive behavior, however, a phase locking mechanism typical of the jet-bore interaction makes the pipe tone perfectly harmonic. Thus, the dispersion, modeled by an all-pass filter, only shifts the resonant peaks of the waveguide, altering the pipe excitation spectrum.}

Of the 58 pipe parameters, 32 are envelope parameters regulating transients times amplitudes, while the remaining part controls static characteristics of the tone. All parameters are time-invariant, thus the evolution of the pipe tone cannot be controlled after the note onset event.

There are dependencies between components of the algorithm  which concurrently shape some of the perceptual features of the sound. As an example, variations of the upper lip of the pipe and of the bore diameter, e.g., have both impact on the harmonic envelope by creating dips and peaks. The first is modeled in the tonal generator as the delay of the comb filter, while the second is modeled as the dispersion filter cutoff frequency in the passive resonator. 

\subsection{Datasets}
\label{subsec:dataset}

\subsubsection{Contrived Dataset}
A dataset of physical modelling stops has been created by sound designers in the previous years {following manual sound design procedures described in Section \ref{subsec:impactsoundesign}}. With such approach the goal was to obtain pleasant sounding stops with resemblance to pipes of a certain historical period. This dataset covers most organ families. We refer to this dataset as the \textit{contrived} dataset, being composed of tones that can be perfectly matched by the physical model.

Each item in the dataset is composed of a parameter set and a 4s-long tone created by the model. Items are labelled according to family, footage and note number (from 0 to 73, corresponding to the range F1-F\#7). The availability of a large dataset allows to split it in subsets that have been used for training the neural networks used in the NS. 


The subsets used for training follow:
\begin{itemize}
\item \textbf{Principale}
\begin{itemize}
\item subset1: Principale 132 stops (\textbf{8'})
\item subset2: Principale 256 stops (\textbf{4'}-\textbf{8'}-\textbf{16'})
\item subset3: Principale 330 stops {(all contrived Principale stops)
\item subset4: Random selection of 90 stops from subset3}
\end{itemize}
\item \textbf{Bordone}
\begin{itemize}
\item subset5: Bordone 56 stops (\textbf{8'})
\item subset6: Bordone 150 stops {(all contrived Bordone stops)}
\end{itemize}
\item \textbf{Flauto Armonico}
\begin{itemize}
\item subset7: Flauto 21 stops {(all contrived Flauto stops)}
\end{itemize}
\end{itemize}

\subsubsection{Non-Contrived Dataset}

A second dataset has been created from non-contrived sounds. This dataset is composed only of recorded tones. Labels associated to each tone are the note number, the family and footage, using the same convention of the contrived dataset. A selection of material has been conducted from dry samples obtained by near-field recording, excluding most of the ambience and reverb. The portion of the dataset used in this work comes from commercial samples taken from the Caen sampleset from Sonus Paradisi \footnote{http://www.sonusparadisi.cz/en/organs/france/caen-st-etienne.html}, covering all the stops of an organ by Aristide Cavaill\`{e}-Coll, one of the most noted organ French Romantic builders. Samples are available from almost all the keys without the use of resampling, thus, providing as a good reference of the original pipe sound. These are used for evaluation in the next sections.

\subsection{Features}
\label{subsection:features}
\begin{figure}[ht!]
\centering
\includegraphics[width=0.99\linewidth]{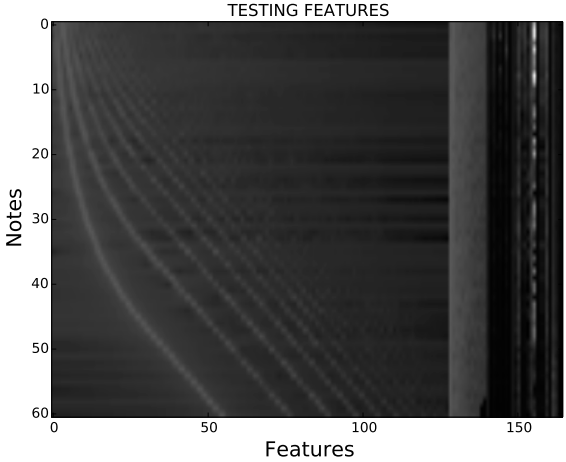}
\caption{2D plot showing features from an entire stop (61 notes). The features are reported along the horizontal axis. The features shown are from left to right: 128 log-Mel coefficients, first 12 harmonics amplitude, SNR, 20 envelope features.}
\label{fig:testingFeatures}
\end{figure}

The feature set employed for the present use case has been devised considering a flue pipe tone as composed of a short attack transient and a steady state phase. The steady state of pipes is known to be perfectly harmonic. Following this knowledge, a feature set has been crafted that contains spectral information, attack envelope information and coefficients related to the noise. Table \ref{tab:Features} reports  further details. The amplitude of the harmonics are meant to describe the steady state periodic spectrum, while the SNR describes the ratio between the latter and the stochastic component of the steady state spectrum. The attack and sustain coefficients track the tone envelope, a useful information for the Neural Network to estimate the envelope parameters of the physical model. Additionally, Logmel coefficients have been added to {test their performance in conjunction with the other features or alone}.
Figure \ref{fig:testingFeatures} shows a feature plot where the leftmost part is composed by the logmel coefficients.

\begin{table}[ht]
	\centering
	\caption{Features used in MLP neural network training}
	\label{tab:Features}
		\resizebox{.5\textwidth}{!}{\begin{tabular}{|c|c|c|}
		\hline
			\multirow{2}{*}{\textbf{Feature Name}}&\textbf{Feature Length}&\multirow{2}{*}{\textbf{Description}}\\
			&$[min,max]$ & \\
			\hline
			\multirow{2}{*}{Harmonics}  &\multirow{2}{*}{$[10,100]$} & Harmonics amplitudes \\
			 & & from steady state FFT\\
			\hline
			\multirow{2}{*}{SNR}&\multirow{2}{*}{$[1]$} & Periodic to \\
			& & stochastic power ratio\\
			\hline
			Logmel &$[64,256]$ & Logmel Spectrum\\
			\hline
			\multirow{3}{*}{Attack\&Sustain }& \multirow{3}{*}{$[12,24]$} & Attack and sustain \\
			&& characterization\\
			&& of first P harmonics\\
			\hline
		\end{tabular}}
\end{table}

\subsection{Objective Metrics}
\label{subsec:obj-metrics}
The use of acoustic metrics is at the core of the SS and the MORIS stages. These are specific for the timbre matching task and may be orthogonal, as is the case with metrics that describe temporal or frequency envelopes, or may concur in describing and giving different weight to similar aspects. The weighting given to each metric concurring to the cost function is arbitrary and must be guided by the sound designer. In the current use case the sound is independent of dynamic control (e.g. keyboard velocity), allowing to adopt orthogonal metrics for the spectral content and the envelope. 

In the following all the metrics used are listed.

\subsubsection{Harmonic Mean Squared Distance Function}
Mean squared distance calculated between harmonics of two periodic signals of same pitch. It measures the difference of each pair of isofrequential harmonics, up to the $L$-th ones, measured from the DFT magnitude spectrum $S_1(\omega)$ and $S_2(\omega)$ computed on the steady state portion of signals $s_1[n]$ and $s_2[n]$.
\begin{equation}
H_L = \frac{1}{L} \sum_{l=1}^{L}  (S_1(l\omega_0) - S_2(l\omega_0))^2
\end{equation}
Along the paper the distance will be denoted as $H_H$ when it is evaluated up to the highest harmonic below the Nyquist frequency.

A weighted harmonic mean squared distance has been also used. This version weights each difference by the amplitude of the harmonic in the target timbre, thus, reducing the importance of the distance when the harmonic to match is low or nearly imperceptible.

\begin{equation}
H_L^W = \frac{1}{L} \sum_{l=1}^{L}  (S_1(l\omega_0) - S_2(l\omega_0))^2S_1(l\omega_0)
\end{equation}

\subsubsection{Envelope Distance Function}
The envelope distance function measures the difference between two envelopes. In the present case it is evaluated only for the attack transient. It is evaluated as the squared difference between two envelopes calculated using the Hilbert transform, i.e.:
\begin{equation}
E_D = \sum_{n=0}^{T_s}{ ( |\mathcal{H}(s_1[n])| - |\mathcal{H}(s_2[n])| )^2 }
\end{equation}
Where $T_s$ is the end of the attack transient.

Additionally, the same metric can be applied separately to a single harmonic, {extracted from the signal by a narrow bandpass filter. Since the flue pipe model of Section \ref{subsec:fluemodel} has independent envelope control of the first and second harmonic, in the present work we focus on the metrics $E_{D1}$, $E_{D2}$, respectively.}

\section{Experiments and Results}
\label{sec:results}

This section provides implementation details related to the experimental setting.
Results are reported exploring both objective and subjective evaluations. For objective evaluations results will be presented employing harmonic mean square distances and the envelope distances, comparing the results with previous works.
Subjective tests have been conducted following an approach similar to MUSHRA \cite{MUSHRA1534-3-2015} on 14 subjects with different musical backgrounds.

\subsection{Implementation details}
\label{subsec:impl-details}

The framework has been implemented in the Python language employing Keras\footnote{http://keras.io} libraries and Theano\footnote{http://deeplearning.net/software/theano/} as a backend, running on a Intel i7 Linux machine equipped with 2 x GTX 970 graphic processing units for neural networks training. The MORIS runs on the same machine but on CPUs. The physical model is implemented as a C++ Linux binary running from console.

A brief overview on training, testing and optimization times is provided.
Convolutional Neural Networks (CNN) training and testing times are reported for comparison.
\begin{itemize}
\item \textit{CNN Training:} \textbf{30-300 [s]} per epoch (depending on the dataset and the network size and parameters).
\item \textit{CNN Testing:}\textbf{ 10-30 [s]} (depending on the network size).\\

\item \textit{MLP Training:} \textbf{\textless20 [s]} per epoch (depending on dataset and network).
\item \textit{MLP Testing:} \textbf{\textless10 [s]} (depending on network).\\
\item \textit{Selection Algorithm (SA):} \textbf{\textless5 [s]} per comparison. Selection Algorithm runs on CPU.\\

\item \textit{MORIS:} \textbf{3-12 [s]} per Iteration. For every note in iteration a 4 seconds long waveform is synthesized to obtain a set of notes composed of both stationary and transitory state. Times can be reduced generating shorter notes.
\end{itemize}

All model parameters in the training set are normalized to the range [-1, 1], parameter-wise.
The cost function for the training is the mean squared error (MSE), minimized using alternatively one of Stochastic Gradient Descent (SGD) \cite{Rumelhart86-LRB}, Adam or Adamax \cite{kingma2014adam} optimization algorithms.
Parameters exploration ranges are reported in table \ref{tab:hyperparamsMLP}.
Batch normalization \cite{ioffe2015batch} and dropout \cite{srivastava2014dropout}  were explored.
\begin{table}[ht]
	\centering
	\caption{Hyperparameters ranges used for the MLP experiments of this work.}
	\label{tab:hyperparamsMLP}
		\resizebox{.5\textwidth}{!}{\begin{tabular}{|c|c|c|}
			\hline
			\textbf{Network}  \rule{0pt}{8pt}	&  \textbf{Layers} &  \multirow{2}{*}{\textbf{Activations}} \\
			\textbf{layout}	 	&  \textbf{sizes} & \\
			\hline
			
			Fully Connected layers: & \multirow{2}{*}{$2^i, 5\leq i \leq12$ }& \multirow{2}{*}{\textit{tanh} or \textit{ReLU}} \\
			2, 3, 4,...,12\rule{0pt}{8pt} &&\\
									 
			\hline
			\textbf{Training }  \rule{0pt}{8pt} & \textbf{Batch} & \textbf{Optimizer parameters} \\ 
			\textbf{epochs}	& 	\textbf{size} &	 \textbf{(SGD, Adam, Adamax)}\\ 
			\hline
			4000, 400 patience  \rule{0pt}{8pt} & 10 to & learning rate = $10^i, -8 \leq i \leq-2$ \\
			Validation split = 10\% &  2000 &  MomentumMax = 0.8, 0.9 \\
			\hline
		\end{tabular}}
\end{table} 

A set of 210 Neural Networks were trained with different combinations of hyper-parameters in supervised random configuration. Selected Neural Networks were chosen based on the Mean Absolute Error (MAE) achieved on a target testing set. Every Neural Network is trained on a subset of contrived sounds as reported in Section \ref{subsec:dataset}. 

For the Heuristic Search Refinement Algorithm the optimization lasts for 4000 iterations.

\subsection{Objective Results}
\label{subsec:obj-results}
This section presents results of the proposed method in acoustic terms employing different metrics. Evaluations are computed and reported separately for each step, from the NS alone to the cascade of NS and SS and finally of all three stages. {Comparison with the previous method is also reported for some of the combinations, showing significantly lower results.}

Target stops are both from the contrived and the non-contrived datasets and do not overlap with those used during training of the Neural Networks.
Results are given on \textbf{8'} stops with fundamental frequency spanning from 43.7~Hz to 2960~Hz. Results are reported for two contrived Principale stops, (\textit{Principale\_VS, Principale\_Stentor\_IT}), two non-contrived Principale stops, (\textit{Principal\_C\_CAEN, Salicional\_C\_CAEN}), one contrived Bordone stop (\textit{Bourdon\_G\_OS\_FR}) and one non-contrived Bordone stop (\textit{CorDeNuit\_C\_CAEN}), one contrived Flauto Armonico stop (\textit{FluteHarmonique\_VS}) and one non-contrived Flauto Armonico stop (\textit{FluteHarmonique\_G\_CAEN}).

\begin{figure}[tbp]
 \centering
 \begin{subfigure}[b]{0.48\textwidth}
 	\includegraphics[width=\textwidth]{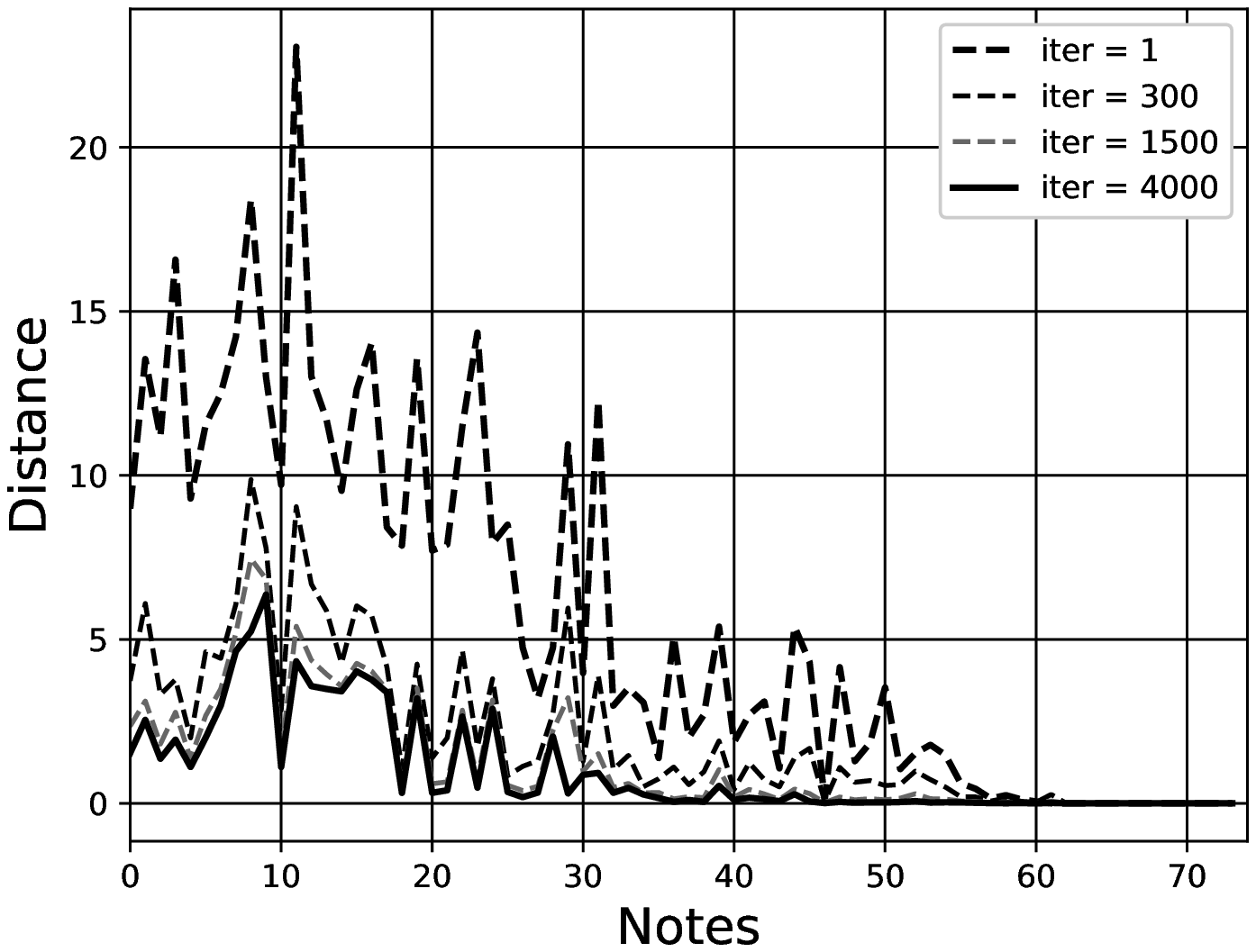}
 	\subcaption{}
 \end{subfigure}
 
 \begin{subfigure}[b]{0.48\textwidth}
 	\includegraphics[width=\textwidth]{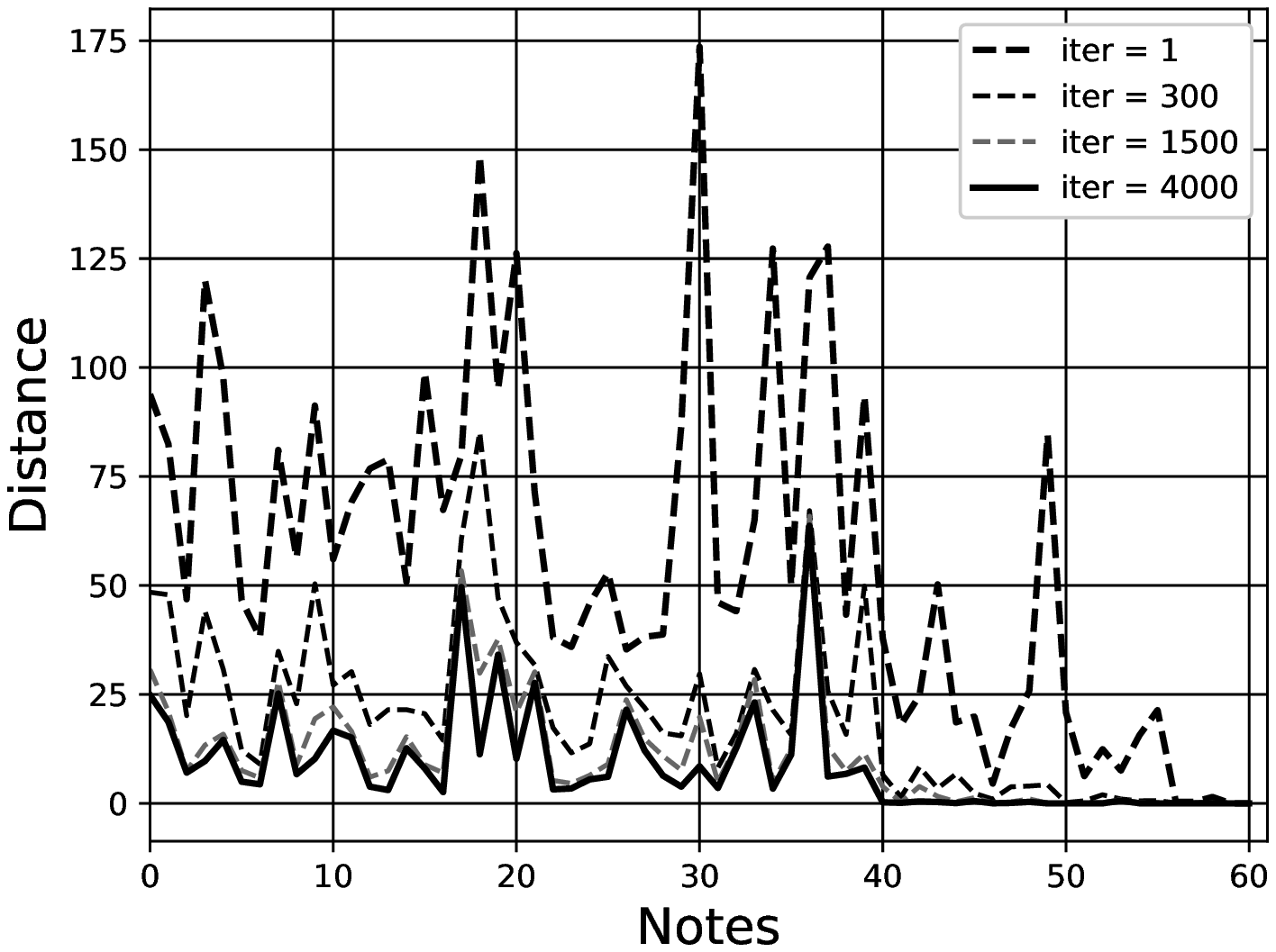}
 	\subcaption{}
 \end{subfigure}
 
 \caption{The reduction of the acoustic distance in the MORIS process for a contrived (a) and non-contrived (b) stop. The distance is calculated from H, $H_{10}$ and $H_{10}^W$ metrics using weights respectively of $[1, 1, 3]$.The distance shown is measured at the beginning of the process and after 300, 1500 and 4000 iterations. The cost function to minimize is the sum of H, $H_{10}$ and $H_{10}^W$, shown for each one of the notes of a stop (horizontal axis). For the contrived stop the note range is F1-F\#7, while for the non-contrived stop the note range is C2-C7.}
 \label{fig:metricsEvolution}
 \end{figure}

Figure \ref{fig:metricsEvolution} shows the reduction of the acoustic distance during MORIS optimization for a contrived and a non-contrived stop. With a reduction of the distance between target and matched tones, the improvement gets reduced. This is due to the error-weighting done by the MORIS algorithm. As shown, the reduction of the distance can be of an one order of magnitude.
 
{Table \ref{tab:testsMLP} reports results in terms of three chosen harmonic distances: $H_H$ (up to the Nyquist limit), $H_{10}$ (first ten harmonics) and $H_{10}^W$ (weighted by the target amplitudes), while Table \ref{tab:ResultsEnveleopes} reports the results in terms of the envelope metrics.} Tests NS, SS and MORIS are reported, with NS reporting the results of the best single Neural Network, SS reporting the outcome of the Selection Stage, thus, exploiting all Neural Networks. MORIS is the outcome of the Multi-Objective Random Iterative Search. Please note that two runs are performed, a first MORIS run minimizing harmonics distances and a second MORIS minimizing the envelope distances on the first and second harmonics and on the whole waveform. As expected, convergence with MORIS is reached in shorter time if the previous stages are able to provide a good estimation of the target tones. 

In most of the presented results there is a remarkable improvement from NS to SS output. In those cases where the improvement is little, e.g. with contrived Principale stops, the estimation provided by a single Neural Network was able to provide results above the average. MORIS outcome gives best results for all tests, in many cases providing a large distance reduction from SS. For what concerns the envelope distance an important improvement is obtained using a MORIS with envelope metrics as a cost function. From informal listening tests the last stage seems to allow, for contrived stops, to reach a nearly perfect psychoacoustic matching of the timbre, with all harmonic distances reaching 1~dB or below.

For Bordone stops, the larger errors can be motivated by their distinctive properties, i.e. the lack of even harmonics and the low number of harmonics, thus making the distance estimate altered by large random differences in the spectral regions between odd harmonics or above the last harmonic. Even though a custom metric (e.g. based only on odd harmonics up to the last one present) would produce better results, we decided to keep the same metric for coherence with the other stops.


As a final remark, contrived tones score better than non-contrived ones, as expected. Timbre matching for contrived tones is nearly optimal, suggesting that the reduced scores obtained with non-contrived stops are due to the physical model limitations. In the authors' opinion this highlights another valuable contribution of the proposed method: not only it can be used for the timbre matching, but also to spot and analyze deficiencies of the physical model under exam, by analysis of the differences between the best match obtained with the proposed approach and the target tone.

%

\begin{table*}[htbp]
\centering
\begin{tabular}{|c|c|c|c|c|c|c|c|}
\hline \multicolumn{4}{|c|}{\textbf{Contrived}}&\multicolumn{4}{|c|}{\textbf{Non-Contrived}}\\
\hline
\textbf{}& $\mathbf{H_H}$ & $\mathbf{H_{10}}$ & $\mathbf{H_{10}^W}$ &\textbf{ }& $\mathbf{H_H}$ & $\mathbf{H_{10}}$ & $\mathbf{H_{10}^W}$\\
\hline
\multicolumn{4}{|c|}{\textit{\textbf{Principale VS 8' [P]}}}&\multicolumn{4}{|c|}{\textit{\textbf{Principal 8' (Caen) [P]}}}\\
\hline

\textbf{\cite{Gabrielli2017TETCI}} & 11.04~dB  & 12.78~dB & 16.95~dB & \textbf{\cite{Gabrielli2017TETCI}} & 32.56~dB & 29.51~dB& 68.53~dB\\

\textbf{NS}  & 4.91~dB & 2.44~dB & 3.36~dB &\textbf{NS} & 25.69~dB & 23.58~dB& 40.94~dB\\

\textbf{SS}  & 2.95~dB & 1.06~dB & 1.41~dB &\textbf{SS} & 14.24~dB & 10.14~dB & 21.89~dB\\

\textbf{MORIS}  & \textbf{2.32~dB} & \textbf{1.01~dB} & \textbf{1.00~dB}&\textbf{MORIS} & \textbf{4.58~dB} & \textbf{2.60~dB }& \textbf{3.41~dB}\\
\hline

\multicolumn{4}{|c|}{\textit{\textbf{Principale Stentor 8' [P]}}}&\multicolumn{4}{|c|}{\textit{\textbf{Salicional 8' (Caen) [P]}}}\\
\hline

\textbf{\cite{Gabrielli2017TETCI}}  & 25.13~dB  & 28.47~dB & 52.07~dB & \textbf{\cite{Gabrielli2017TETCI}} & 182.66~dB & 232.84~dB& 1163.57~dB\\

\textbf{NS}  & 3.86~dB & 1.58~dB & 3.31~dB &\textbf{NS} & 17.67~dB & 13.26~dB & 30.20~dB\\

\textbf{SS}  & 3.26~dB & 1.53~dB & 2.36~dB &\textbf{SS} & 11.25~dB & 8.08~dB & 15.56~dB\\

\textbf{MORIS}  &\textbf{1.32~dB} & \textbf{0.52~dB} & \textbf{0.77~dB} &\textbf{MORIS} & \textbf{4.93~dB} & \textbf{3.48~dB} & \textbf{3.87~dB}\\
\hline
%
%
%
\multicolumn{4}{|c|}{\textit{\textbf{Bourdon GO FR 8' [B]}}}&\multicolumn{4}{|c|}{\textit{\textbf{Cor de Nuit 8' (Caen) [B]}}}\\
\hline

\textbf{NS}  & 46.66~dB & 53.31~dB & 115.49~dB &\textbf{NS} & 44.71~dB & 50.16~dB & 88.65~dB\\

\textbf{SS}  & 21.65~dB & 21.11~dB & 24.73~dB &\textbf{SS} & 24.74~dB& 21.21~dB & 32.96~dB\\

\textbf{MORIS}  &\textbf{11.50} & \textbf{8.63~dB} &\textbf{10.18~dB} &\textbf{MORIS} &\textbf{8.84~dB} &\textbf{4.60~dB} & \textbf{4.87~dB}\\
\hline
\multicolumn{4}{|c|}{\textit{\textbf{Flute Harmonique VS 8' [FA]}}}&\multicolumn{4}{|c|}{\textit{\textbf{Flute Harmonique 8' (Caen) [FA]}}}\\
\hline

\textbf{NS}  & 10.90~dB & 5.34~dB & 16.67~dB &\textbf{NS} & 36.68~dB & 56.87~dB & 59.32~dB\\

\textbf{SS}  & 8.58~dB & 2.30~dB & 6.93~dB &\textbf{SS} & 19.91~dB & 13.89~dB & 29.78~dB\\

\textbf{MORIS}  & \textbf{1.40~dB} & \textbf{0.49~dB} & \textbf{0.38~dB} &\textbf{MORIS} & \textbf{15.86~dB} & \textbf{11.85~dB}  & \textbf{12.27~dB}\\
\hline
\end{tabular}
\caption{Timbre matching results reported in terms of harmonic distance at the output of each stage of the proposed approach for stops belonging to the Principale (P), Bordone (B) and Flauto Armonico (FA) subfamilies. Where available, the results from the end-to-end approach of  \cite{Gabrielli2017TETCI} are reported. Please note that NS is the outcome of the best single neural network among all the neural networks used for the subsequente SS. MORIS has been carried on for 4000 iterations.}
\label{tab:testsMLP}
\end{table*}

\begin{table*}[htbp]
\centering
\begin{tabular}{|c|c|c|c|c|c|c|c|}
\hline 
\multicolumn{4}{|c|}{\textbf{Contrived}}&\multicolumn{4}{|c|}{\textbf{Non-Contrived}}\\
\hline
\textbf{}& $\mathbf{E_{D2}}$& $\mathbf{E_{D1}}$ & $\mathbf{E_D}$ &\textbf{}&  $\mathbf{E_{D2}}$& $\mathbf{E_{D1}}$ & $\mathbf{E_D}$\\
\hline
\multicolumn{4}{|c|}{\textit{\textbf{Principale Stentor 8' [P]}}}&\multicolumn{4}{|c|}{\textit{\textbf{Salicional 8' (Caen) [P]}}}\\
\hline

 \textbf{SS}  & 230.47 & 86.71 & 394.18 & \textbf{SS}  & 2053.02  & 6430.48 & 3211.78 \\

\textbf{MORIS}  & \textbf{5.48} & \textbf{9.59} & \textbf{341.12} &\textbf{MORIS} & \textbf{1380.81} & \textbf{1633.63} & \textbf{2326.01}\\
\hline

\end{tabular} 
\caption{Timbre matching results reported in terms of envelope distance at the output of the SS and MORIS stages, for two Principale stops. Results are presented exploiting envelope distances. The MORIS stage has been run for 300 iterations with the $E_D$, $E_{D1}$ and $E_{D2}$ as cost function. Results are presented for two stops only for conciseness, but similar results are found with all other tested stops.}
\label{tab:ResultsEnveleopes}
\end{table*}

\subsection{Subjective Tests}
\label{subsec:subjective-results}
Listening tests have been conducted to assess the effect of the reduced acoustic distances in psychoacoustic terms. The subjective tests are inspired by the MUSHRA method \cite{MUSHRA1534-3-2015}. With MUSHRA a user is exposed to several stimuli, and he/she must rate the similarity of each of them with respect to a reference tone using a 1-100 scale. The method also requires that at least one \textit{anchor} and a hidden copy of the reference are provided among the stimuli. The mandatory anchor is a tone providing a bottom line to the test, and this is done by degrading its quality using a 3.5~kHz low-pass filtered version of the reference. This fits well the use case of audio coding algorithms or sound reproduction systems, where one of the quality criteria is the bandwidth of the audio output, however in our case this choice is not perfectly viable because many organ tones may have very low energy over 3.5~kHz resulting in a hardly perceptible degradation. The recommendations \cite{MUSHRA1534-3-2015} suggest the addition of other types of anchors providing similar types of impairments as the system under test, e.g. additional noise, packet loss and dropouts in the case of audio transmission systems. Following these guidelines and considering that the kind of impairment provided by a bad timbre matching can be, e.g. the presence of excess wind noise and a mismatch of the harmonic content, we devised an anchor by manually modifying a Principale tone. The noise gain in the physical model was raised in order to have an RMS value that is +30dB with respect to the harmonic component and the first and second harmonics were not clipped.

The stimuli under test consisted in a tone generated by the CNN approach of \cite{Gabrielli2017TETCI} and a tone generated by the output of the proposed algorithm (indicated as PROP) including all stages. To resume, the subjects were exposed for each screen of the test to the reference tone and the following stimuli in random order: a hidden reference tone, the proposed anchor, the CNN estimation and the full NS+SS+MORIS estimation. The test consisted of eight screens with a first warm-up screen, not employed in the evaluation. Four screens were related to a contrived reference tone and four to a non-contrived reference tone. The whole test took 15 minutes per subject on average.
Tests have been conducted on 20 subjects, 15 male and 5 female aged 16-53 with varying musical background. After the tests have been completed 6 of the subjects were discarded, following MUSHRA guidelines, because they were not able to reliably repeat the assessment of the hidden reference, specifically, they rated it for more than 15\% of the test items with a score lower than 90.

Table \ref{tab:Subjective-Tests} reports data for each of the 14 selected subjects, providing the average rating for CNN tones and PROP tones for each subject, the musical background of each subject and the average and standard deviation of the CNN and PROP tones for the whole session of tests. Every single subject evaluated the CNN results worse than PROP results. The average distance is 16.2 and the difference between the two approaches is statistically significant. The significance test have been done considering the null hypothesis $h_0: \mu=\overline{PROP}-\overline{CNN}<0$ with a p-value of $0.1$, where $\overline{CNN}$ and $\overline{PROP}$ are respectively average values of CNN and PROP per subject. 

Statistical test on hypothesis $h_0$ has been conducted exploiting a Student's t-distribution with 14 degrees of freedom. 




\begin{table}[ht!]
\centering
\begin{tabular}{|c|c|c|c|}
\hline \multirow{2}{*}{\textbf{Subject}} & \multirow{2}{*}{$\overline{\textbf{CNN}}$}  & \multirow{2}{*}{$\overline{\textbf{PROP}}$ }&\multirow{2}{*}{\textbf{Exp(Y)}}\\&&&\\

\hline
\textbf{S1} & 40.625 & 66.500 & 12\\

\textbf{S2} & 50.125 & 71.750 & 9\\

\textbf{S3} & 68.125 & 80.000 & 12\\

\textbf{S4} & 58.125 & 75.375 & 0\\

\textbf{S5} & 53.125 & 58.625 & 19\\

\textbf{S6} & 58.500 & 60.250 & 9\\

\textbf{S7} & 51.000 & 64.375 & 30\\

\textbf{S8} & 30.125 & 55.875 & 0\\

\textbf{S9} & 53.000 & 63.375 & 14\\

\textbf{S10} & 32.500 & 63.125 & 0\\

\textbf{S11} & 48.625 & 74.125 & 10\\

\textbf{S12} & 44.769 & 56.538 & 38\\

\textbf{S13} & 45.538 & 66.769 & 40 \\
 
\textbf{S14} & 53.538 & 66.308 & 38\\
\hline
\textbf{Overall Average} & 49.7  & 65.9 & \\
\hline
\textbf{Standard Deviation} & 10.4 & 7.2 &\\
\hline
\textbf{Average Difference} & \multicolumn{2}{c|}{\textbf{16.2}} &\\
\hline
\end{tabular}
\caption{Subjective test summary table. Per-subject and averaged test values are reported. The last column reports the musical experience for each subject expressed as the number of years of study and practice.}
\label{tab:Subjective-Tests}
\end{table}

\subsection{Impact on the Sound Design Process}\label{subsec:impactsoundesign}
{After introducing the new sound design method, an informal evaluation has been conducted to shed light on the impact of the new approach. We report some qualitative information gathered during informal discussions with the sound design team that developed the model and the sound libraries commercially available for the aforementioned physical model since year 2004.}

Before the introduction of the proposed approach, the sound design process was completely manual, based on the knowledge and the skills of the sound designers and never had the goal of matching a set of samples. Therefore, the new approach allows to set new goals for the sound design team.

The manual approach that had been used previously followed these steps: 
\begin{itemize}
\item select a reference stop family of a specific style, related to either historical or geographical factors influencing the target timbre;
\item start from a template of parameters or a similar stop if already designed;
\item manually interact with a graphical software to alter the parameters of a note based on the knowledge of the physical model and the causal relation between parameters and sound until a desired timbre is obtained;
\item repeat previous step for a subset of notes;
\item interpolate between the selected notes to obtain values for the rest of the notes in the keyboard range, eventually adding some random fluctuations.
\end{itemize}

Please note that the reference stop may not be an existing stop, or a stop available to the sound designer in form of recorded tones, but it may be a mental representation of a prototype sound from a specific style or historical organ builder.


Development times for such an approach are strongly dependent on human factors such as fatigue, experience, knowledge of the physical model and efficiency of the graphical software interface. From the informal discussion it seems that preparing a complete stop required an effort that is of the order of magnitude of a working day (8 hours) split in several sessions to recover from fatigue, while with the proposed approach and the current computational resources one stop can be prepared in approximately 5 minutes to which some human effort must be added to judge on the result and conduct some final adjustments. This is an advancement compared to the findings in \cite{Tatar2016OP1} and allows a fast paced generation of sound libraries.

To conclude, it must be noted that the computational process allows for an objective result, while, previously, target tones were seldom used, as the matching would hardly be feasible in a reasonable amount of time. As a downside, the objective approach requires data, specifically, good quality recordings of single tones, not always available at ease. It is worth noting that the expertise of the sound designer is still very valuable even with the proposed timbre matching approach as he/she is still responsible for all the design choices and the supervision of the computational work.

\section{Conclusions}
\label{sec:conclusion}
In this paper a novel multi-stage estimation algorithm is presented and applied to the physical model parameters estimation task. The first stage consists in a refinement of previous deep learning techniques, where feature learning is replaced by hand-crafted features. Hyperparameter diversity is employed in the subsequent selection stage by picking the best note estimate from a number of differently trained and crafted neural networks. Finally results are refined using an optimization algorithm that employs acoustic distances as cost functions. The first two stages perform a global search in the physical model parameter space, while the last stage performs a local search that converges to a suboptimal solution. 

The algorithm has been validated on a flue pipe physical model as in previous works. The approach is tested on contrived and non-contrived sounds, showing that a large improvement with respect to recent techniques is achieved in both cases. Listening tests have been performed to assess the improvement given by the proposed algorithm with respect to the previous end-to-end algorithm, showing a clear preference of the subjects for the timbre matching provided by the proposed algorithm.

An argument can be made that, when the matching of contrived tones is nearly optimal, the approach can highlight the limitations of the physical model, thus, providing useful insights for improving the model. {The black-box approach allows to generalize its application to any model, by modifying the number of neurons of the last fully connected layer in the neural networks and by creating a specific dataset for the training. This can be done by generating random samples from random sets of parameters as shown in \cite{Gabrielli2017TETCI}. Each use case will require a hyperparameter search to reach optimal results. More experiments to be conducted on open-sourced physical models are on the way.

Some additional effort is still required to allow a comparison between our approach and the genetic algorithms seen in  \cite{Mitchell2007evolutionary,Mitchell2012evolutionary,Tatar2016OP1}. Furthermore, these works employ synthesizers as a use case. It is currently unknown what synthesis engine would be the best for benchmarking the accuracy of parameter estimation algorithms.}

As a last remark, the change from the learned features used in \cite{Gabrielli2017TETCI} to the hand-crafted features is able itself to largely improve the estimation performance. This result may be surprising, given the current trends towards autonomous feature learning in the computational audio processing field. However, other authors have argued that the 2D convolutional approach, as developed by image processing researchers, is not adequate for audio processing as it does not account for time-frequency anisotropies, and, thus, several modifications of the convolutional approach have been already proposed both on 2D and 1D audio representations \cite{Pons2017audioCNN,Pons2017timbre,Lee2017samplelevelDCNN,Wavenet2016}. More investigation in the machine listening field is required to understand how an accurate computational auditory model can be built to be exploited in the timbre matching task and supersede hand-crafted features.

\ifCLASSOPTIONcaptionsoff
  \newpage
\fi



%

\bibliographystyle{IEEEbib}
\bibliography{biblioCumulativaLeo,bibLeo-ListPub}

\newpage

\begin{figure}[tbp]
\centering
\begin{subfigure}[b]{0.47\textwidth}
	\includegraphics[width=\textwidth]{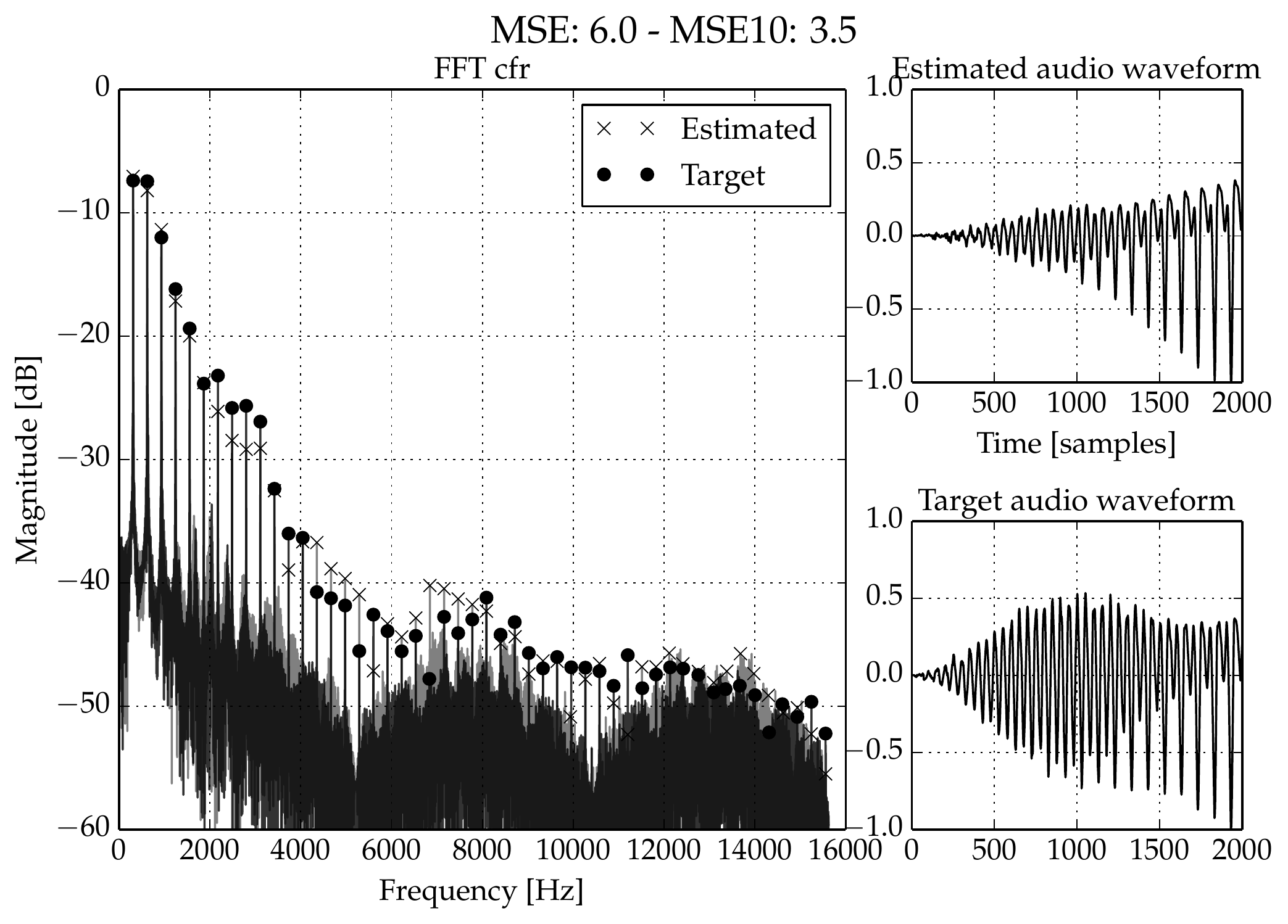}
	\subcaption{}
\end{subfigure}

\begin{subfigure}[b]{0.47\textwidth}
	\includegraphics[width=\textwidth]{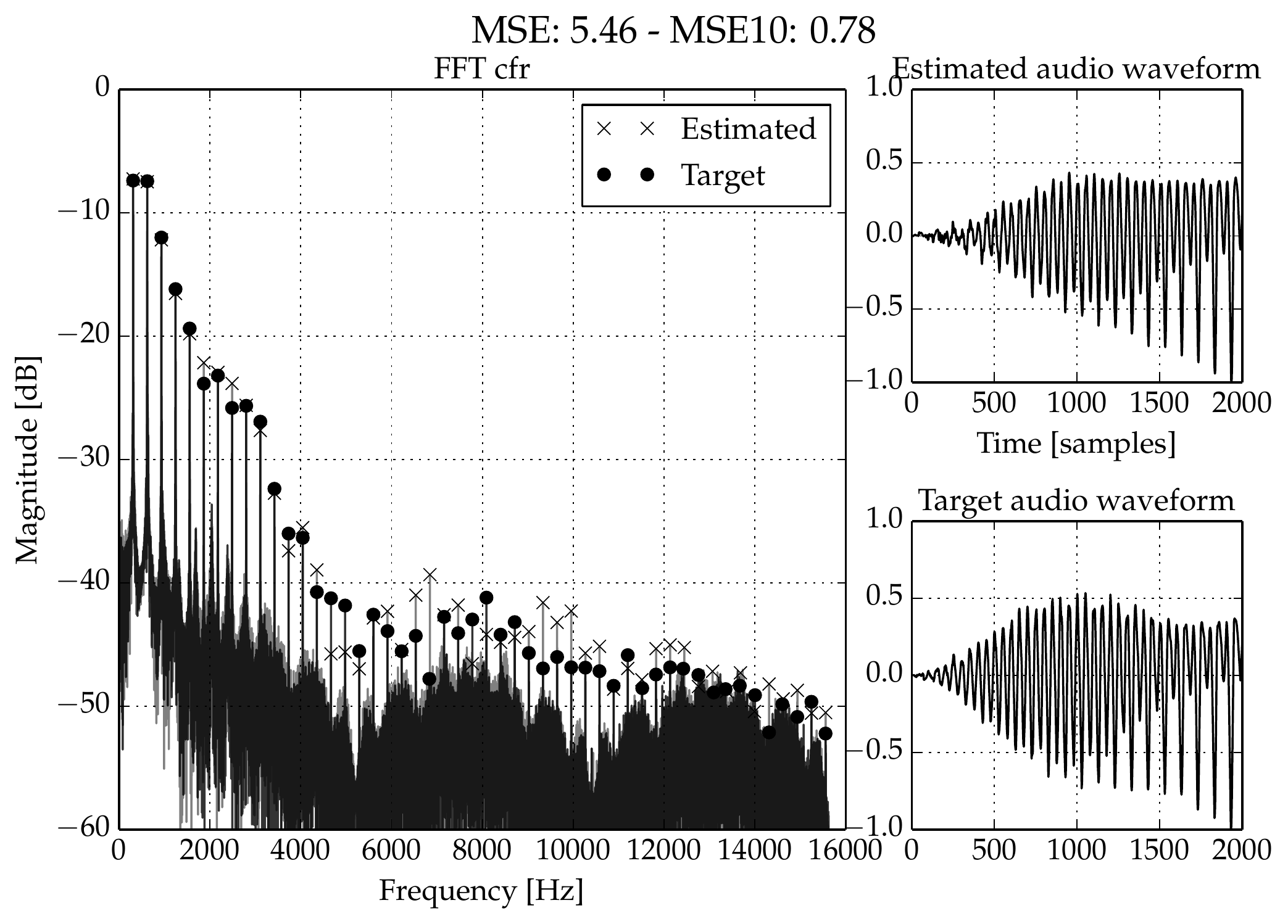}
	\subcaption{}
\end{subfigure}

\begin{subfigure}[b]{0.47\textwidth}
	\includegraphics[width=\textwidth]{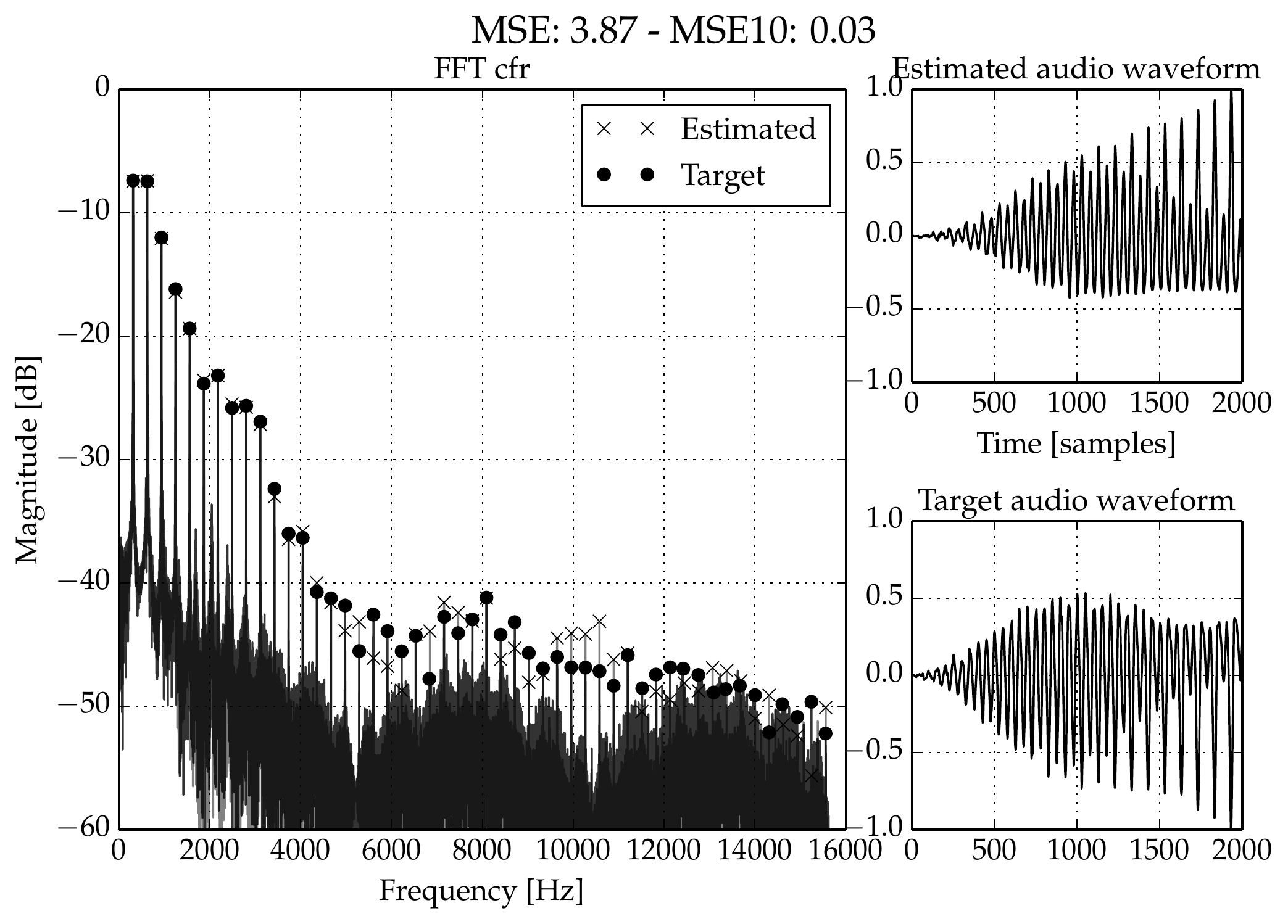}
	\subcaption{}
\end{subfigure}

\caption{Spectra and attack transient in the time domain for contrived \textit{Principale\_VS} D\#4 tone (a) from SS, (b) from NS and (c) from MORIS. H and $H_{10}$ are respectively 6.00~dB, 3.50~dB (a), 5.46~dB, 0.78~dB for (b) and 3.87~dB and 0.03~dB for (c).}
\label{fig:comparison1}
\end{figure}

\begin{figure}[tbp]
\centering
\begin{subfigure}[b]{0.47\textwidth}
	\includegraphics[width=\textwidth]{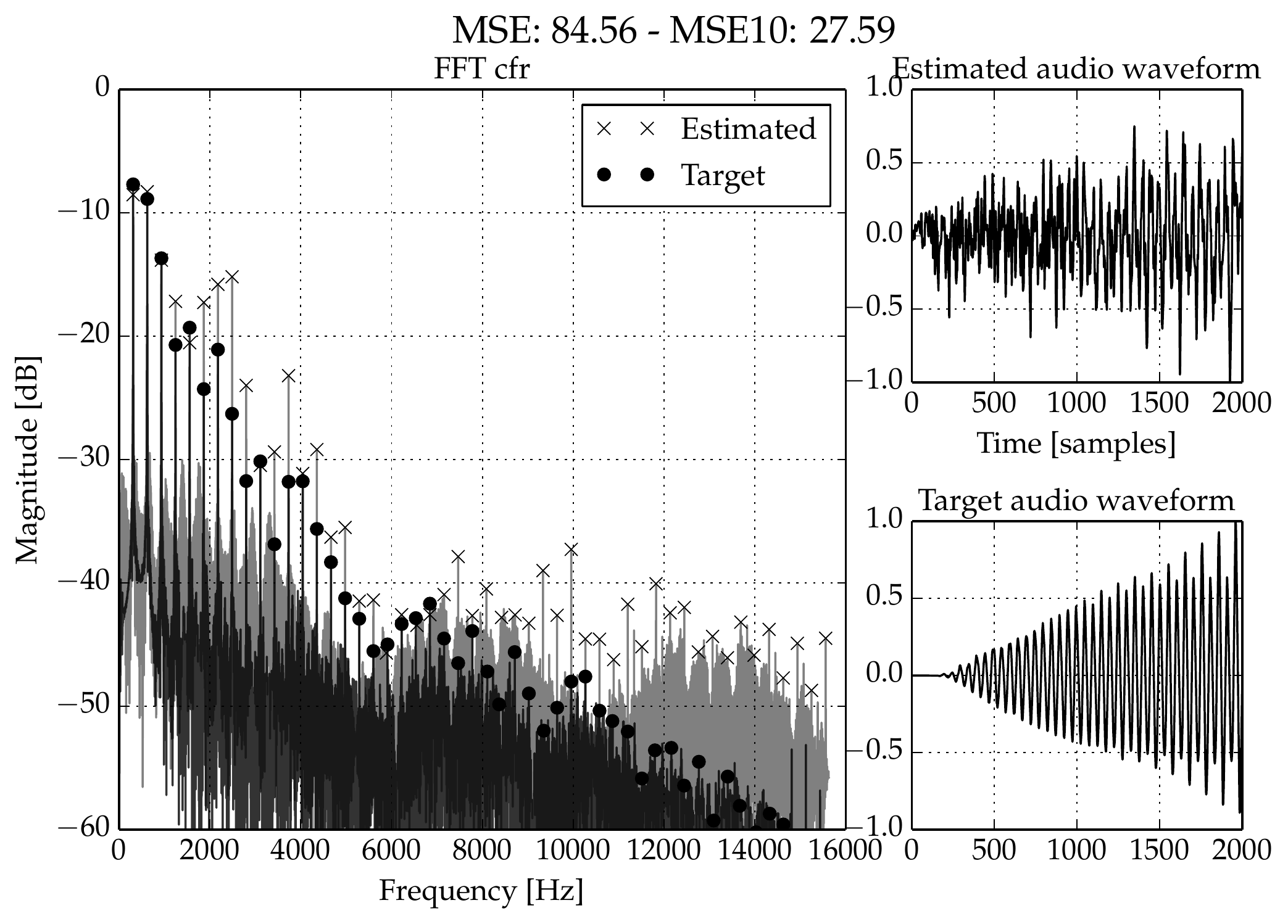}
	\subcaption{}
\end{subfigure}

\begin{subfigure}[b]{0.47\textwidth}
	\includegraphics[width=\textwidth]{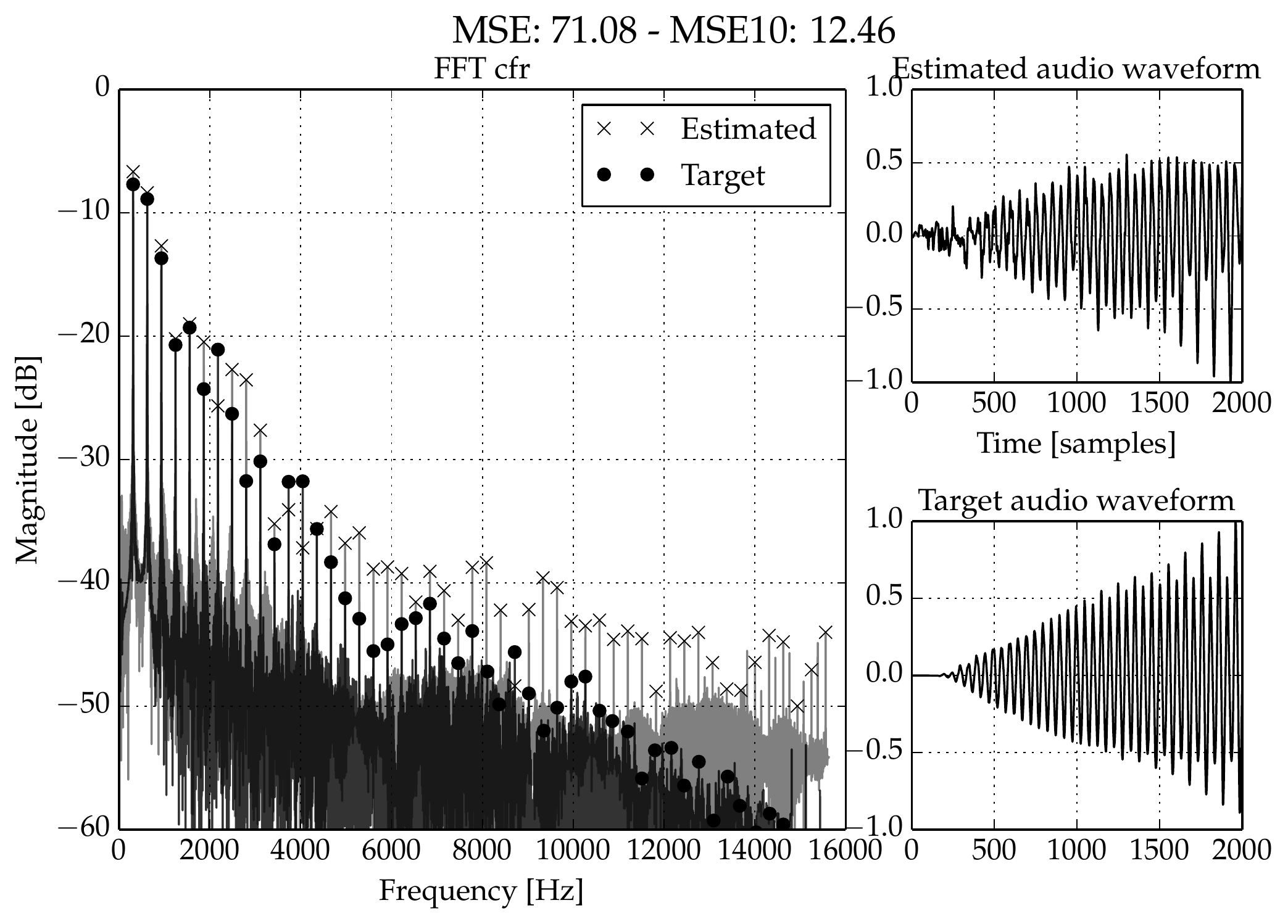}
	\subcaption{}
\end{subfigure}

\begin{subfigure}[b]{0.47\textwidth}
	\includegraphics[width=\textwidth]{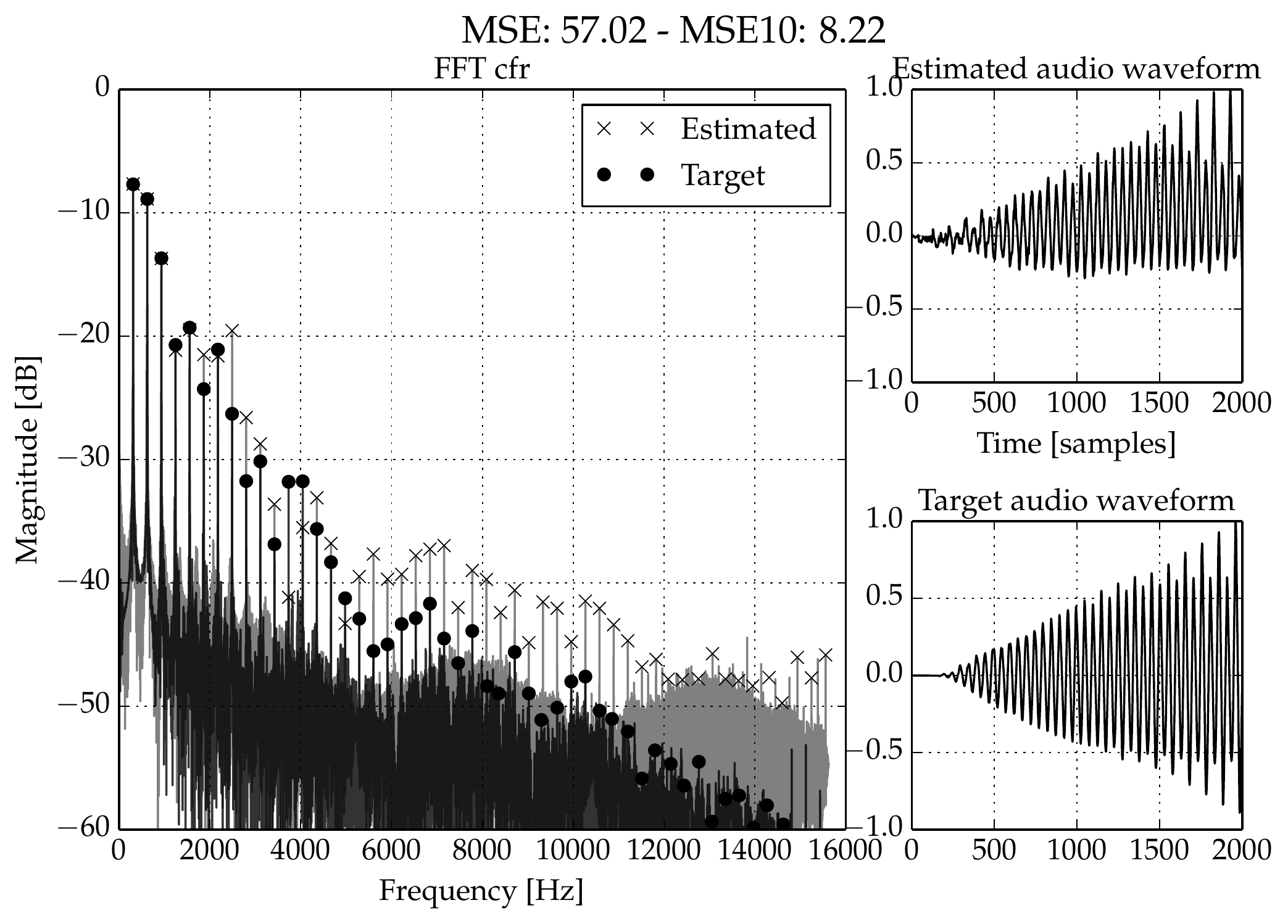}
	\subcaption{}
\end{subfigure}

\caption{Spectra and attack transient in the time domain for non-contrived \textit{Salicional\_C\_CAEN} D\#4 tone (a) from SS, (b) from NS and (c) from MORIS. H and $H_{10}$ are respectively 84.56~dB, 27.59~dB (a), 71.08~dB, 12.46~dB for (b) and 57.02~dB and 8.22~dB for (c).}
\label{fig:comparison2}
\end{figure}

\end{document}